\documentclass{article}
\usepackage{spconf,amsmath,graphicx}
\DeclareMathOperator*{\argmin}{argmin}

\usepackage{enumitem}
\setlist{nosep, leftmargin=14pt}

\usepackage{mwe} 

\title{To what extent can Plug-and-Play methods outperform neural networks alone in low-dose CT reconstruction}
\name{Qifan Xu, Qihui Lyu, Dan Ruan, and Ke Sheng}
\address{University of California, Los Angeles}

\begin{document}
%
\maketitle
\begin{abstract}
The Plug-and-Play (PnP) framework was recently introduced for low-dose CT reconstruction to leverage the interpretability and the flexibility of model-based methods to incorporate various plugins, such as trained deep learning (DL) neural networks. However, the benefits of PnP vs. state-of-the-art DL methods have not been clearly demonstrated. In this work, we proposed an improved PnP framework to address the previous limitations and develop clinical-relevant segmentation metrics for quantitative result assessment. Compared with the DL alone methods, our proposed PnP framework was slightly inferior in MSE and PSNR. However, the power spectrum of the resulting images better matched that of full-dose images than that of DL denoised images. The resulting images supported higher accuracy in airway segmentation than DL denoised images for all the ten patients in the test set, more substantially on the airways with a cross-section smaller than 0.61cm$^2$, and outperformed the DL denoised images for 45 out of 50 lung lobes in lobar segmentation. Our PnP method proved to be significantly better at preserving the image texture, which translated to task-specific benefits in automated structure segmentation and detection.
\end{abstract}
\begin{keywords}
Low-dose CT, Plug-and-Play, deep learning
\end{keywords}
\section{Introduction}
\label{sec:intro}

X-ray CT is instrumental in cancer screening and other medical applications \cite{applications}. However, excessive exposure to X-ray is a health risk factor \cite{risks} and efforts have been made to reduce imaging dose. Two natural ways to do so on commercial CT scanners are 1) to decrease the working current of X-ray tube (LdCT) \cite{LdCT}, and 2) to decrease the number of views per scan \cite{sparse}. These methods unavoidably degrade image quality when standard filtered back-projection (FBP) is used for CT reconstruction. To fully utilize the information in CT sinograms, iterative methods can be used to minimize a regularized cost:
\begin{equation}
\label{eqn1}
    E(x)=\left\lVert Ax-y\right\rVert+\lambda R(x),
\end{equation}
where $x$ is the linear attenuation matrix we aim to estimate, $A$ is the system matrix, $y$ is the measured sinogram, $\left\lVert \cdot \right\rVert$ is the fidelity term encouraging the estimated forward projection to match the measured sinogram, $R(x)$ is the regularization term, and $\lambda$ is the relative weight.

Plug-and-Play (PnP) \cite{zhang2019deep, manifold} is to substitute a certain update in an iteration with a denoiser. Its versatility lends itself to incorporating filters and models, including deep learning (DL) neural networks. Though powerful, DL neural networks' black-box nature prevents straightforward interpretation of the results, a concern particularly relevant to LdCT screening tasks. Therefore, it is of great interest to integrate DL into PnP for both strong statistical learning of latent image information and the rigor of regularized reconstruction.

Previous studies on PnP leveraged frameworks such as consensus equilibrium (CE) \cite{CE, distributedCE}, RED by Romano $et$ $al.$ \cite{RED}, and most frequently, ADMM \cite{ADMM1, airnet, ADMMscalable}. On the other hand, the comparison between PnP with the DL plugin vs. DL alone was inconclusive. Our study aims to elucidate the advantage of PnP incorporating DL vs. DL alone in a rigorously controlled experiment, with a focus on
  \begin{enumerate}
      \item Hyperparameter tuning. As iterative methods, the performance depends heavily on the tuning of hyperparameters such as step size and denoising strength. Hand-tuning may lead to suboptimal results.
      \item DL neural network plugin compatibility. DL neural networks' performance tends to degrade when applied to a dataset with a shifted domain. However, previous works failed to align the domains on which the DL plugins were trained and applied.
      \item Previous works showed improvements on PSNR, MSE and/or other quantitative assessment. However, the standard image quality metrics can be insensitive to the intended tasks of LdCT for fine feature detection and delineation.
  \end{enumerate}

\section{Methods and materials}
\label{sec:methods}
\subsection{Overall framework}
Among the frameworks employed by previous works such as ADMM, CE, RED, and projected gradient descent \cite{projected}, we employed half quadratic splitting \cite{zhang2019deep}. The objective function and the optimization iterations are:
\begin{equation}
    \hat x, \hat v=\argmin_{x, v}\frac{1}{2\sigma^2}\left\lVert Ax-y\right\rVert_2^2 + R(v) + \frac{\mu}{2}\left\lVert x-v\right\rVert_2^2,
\end{equation}
\begin{equation}
\label{eqn3}
    v_{k+1}=\argmin_v R(v)+\frac{\mu}{2}\left\lVert x_k-v\right\rVert_2^2,
\end{equation}
\begin{equation}
\label{eqn4}
    x_{k+1}=\argmin_x\frac{1}{2\sigma^2}\left\lVert Ax-y\right\rVert_2^2+\frac{\mu}{2}\left\lVert x-v_k\right\rVert_2^2,
\end{equation}
where $A$, $y$, and $R(x)$ are of the same meaning as in Eqn. \ref{eqn1}. $\left\lVert \cdot \right\rVert_2^2$ is the MSE. The linear attenuation matrix variable is split into $x$ and $v$, individually associated with the data fidelity and prior, coupled by an MSE term. Eqn. \ref{eqn3} would then be replaced with a DL plugin.

For LdCT reconstruction, we used the dataset \cite{TCIA} from the cancer image archive (TCIA) provided by Mayo Clinic. This study focused on lung CT data, including 42 scans from Siemens. Poisson noise was added to the projection to simulate LdCT using 10\% of the standard imaging dose. The study only used the sinograms from the dataset, not the vendor reconstructed images. Reference full- and low-dose images were reconstructed using the iterative method with TV regularization. The 42 patients were split into training (n=28), validation (n=4), and testing (n=10), respectively.

\begin{figure}
    \centering
    \includegraphics[width=\linewidth]{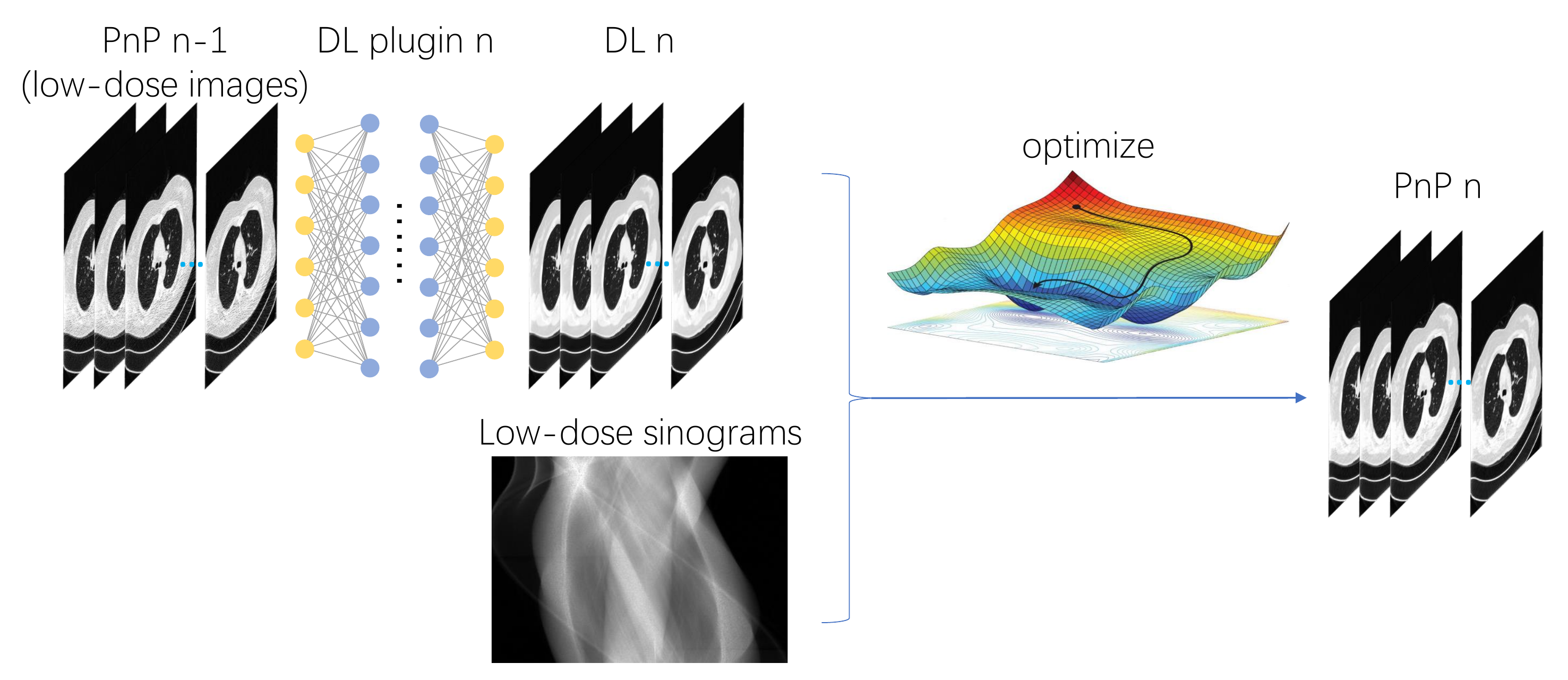}
    \caption{The $n$'th loop for our PnP workflow. The output of the previous loop is taken in as input. For $n=1$, the input is ``low dose images"}
    \label{workflow}
\end{figure}

The wrokflow is composed of a cascade of loops. The $n$'th loop is illustrated in Fig. \ref{workflow}.

\subsection{Network and optimization details}
\begin{figure}
    \centering
    \includegraphics[width=\linewidth]{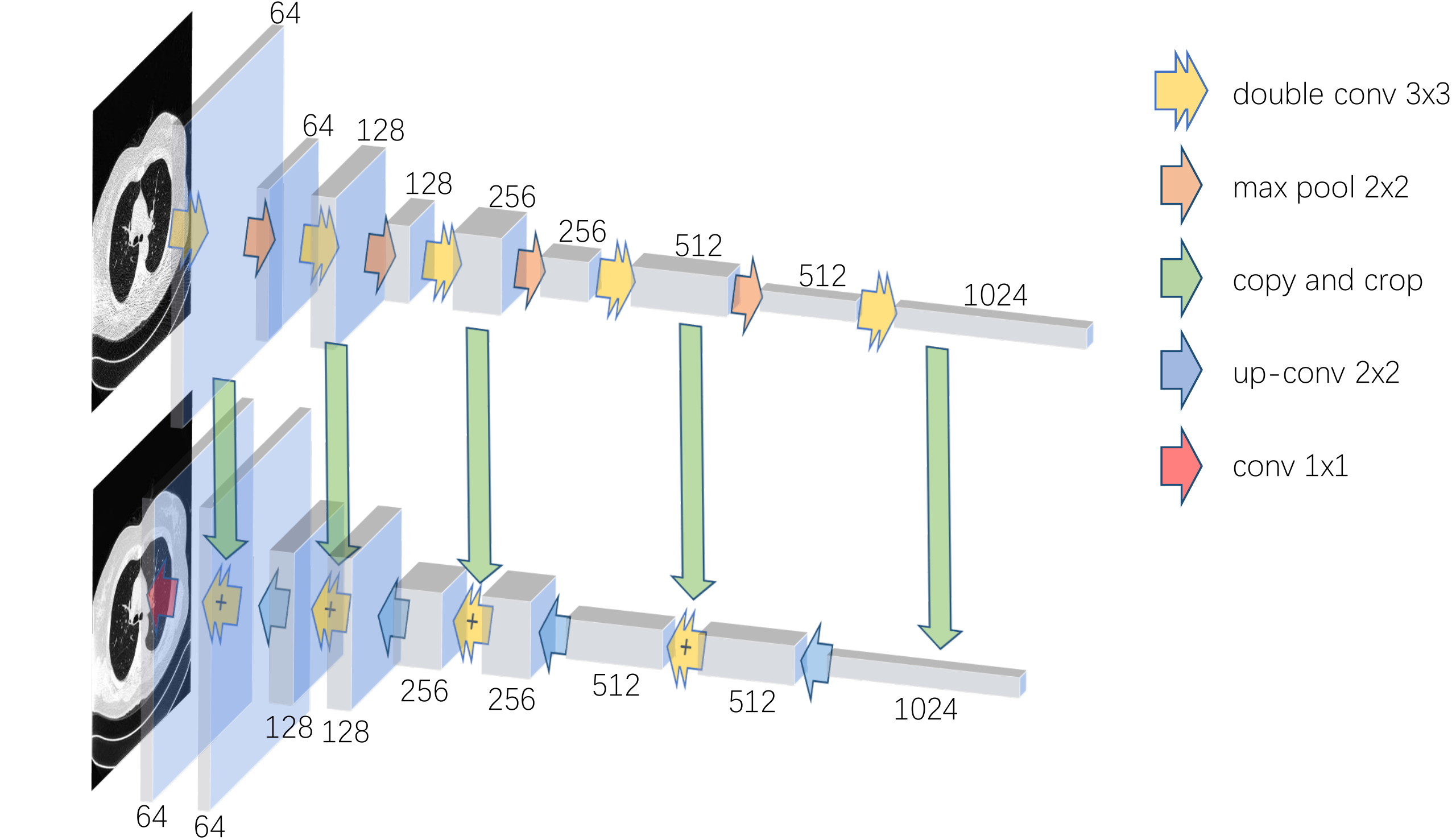}
    \caption{U-Net architecture.}
    \label{UNet}
\end{figure}

We choose to employ U-Net \cite{unet} as our network architecture, as shown in Fig. \ref{UNet}. We subsequently used MSE loss in network training. The trained DL plugin for the first loop was also included in the comparison. We used the conjugate gradient algorithm to optimize the quadratic cost function in Eqn. \ref{eqn4}.

The training phase was implemented in a loop-wise progressive manner. For each loop, we fed ``PnP $n-1$" (or low-dose images for $n=1$) and their corresponding full-dose images into the network, as input and ground-truth, respectively. ``PnP $n$" was obtained after the optimization, which in turn served as the input for the next loop. We ran the workflow in three loops. The parameter $\mu\sigma^2$ of the optimization (Eqn. \ref{eqn4}) was chosen so that the ``PnP 1" had the noise power spectrum most close to that of full-dose images. Specifically, we set it to be 5e3, 7e3, and 8e3 for the three loops, which were then fixed during inference.

\subsection{DL plugin ablation experiment}
To demonstrate the benefits of our loop-specific DL plugin design, we also ran another framework, in which the DL plugins in the second and third loops were replaced with the one of the first loop. We compared the MSE and PSNR values of the DL images of this workflow with those of our proposed framework.

\subsection{GAN denoising network training}
Besides the vanilla ``DL 1" method, to include state-of-the-art DL denoising method in the comparison, we trained a separate denoising U-Net supervised with a hybrid loss as a combination of GAN, SSIM, and L1 losses:
\begin{equation}
    \mathrm{Loss}=\alpha\mathrm{Loss}_{\mathrm{GAN}}+\beta\mathrm{Loss}_{\mathrm{SSIM}}+\gamma\mathrm{Loss}_{\mathrm{L}_1}
\end{equation}

\subsection{Task-relevant image quality metrics}
Metrics such as MSE and PSNR do not reflect spatial characteristics of the texture. For this, we employed noise power spectrum (NPS) \cite{NPS, NPS2} to help evaluate resultant images. For each slice, we obtained the low-frequency components by convolving the original slices with a 4-by-4 averaging convolution kernel. High-frequency components were obtained by subtracting the low-frequency components from the original slices. The intensity of different frequency components for a single slice was computed and averaged over all slices of a patient to determine the patient-specific NPS.

To differentiate texture that was informative for diagnosis from patterned noise, we designed two clinical-relevant tasks, airway and pulmonary lobe segmentation. As we do not have expert annotations for our reconstructed images, we trained automated segmentation networks using well-validated public datasets. The airway dataset \cite{AirwayDataset} contains binary airway annotations for 40 CT scans from LIDC-IDRI \cite{LIDC-IDRI}, and the pulmonary lobe segmentation dataset \cite{PulmonayLobeDataset} contains annotations for 50 CT scans from LUNA16 \cite{LUNA16} dataset. We again used a 2D U-Net, but with eleven consecutive slices for input. For airway segmentation, due to the imbalance between the airway and background, we chose to use Dice's coefficient as the loss function. For pulmonary lobe segmentation, we used cross entropy to stabilize training. The task-specific evaluation was then performed by comparing the automated segmentation on LdCT and full-dose CT images using Dice's coefficient.

\section{Results}
\begin{figure}
    \centering
    \includegraphics[width=\linewidth]{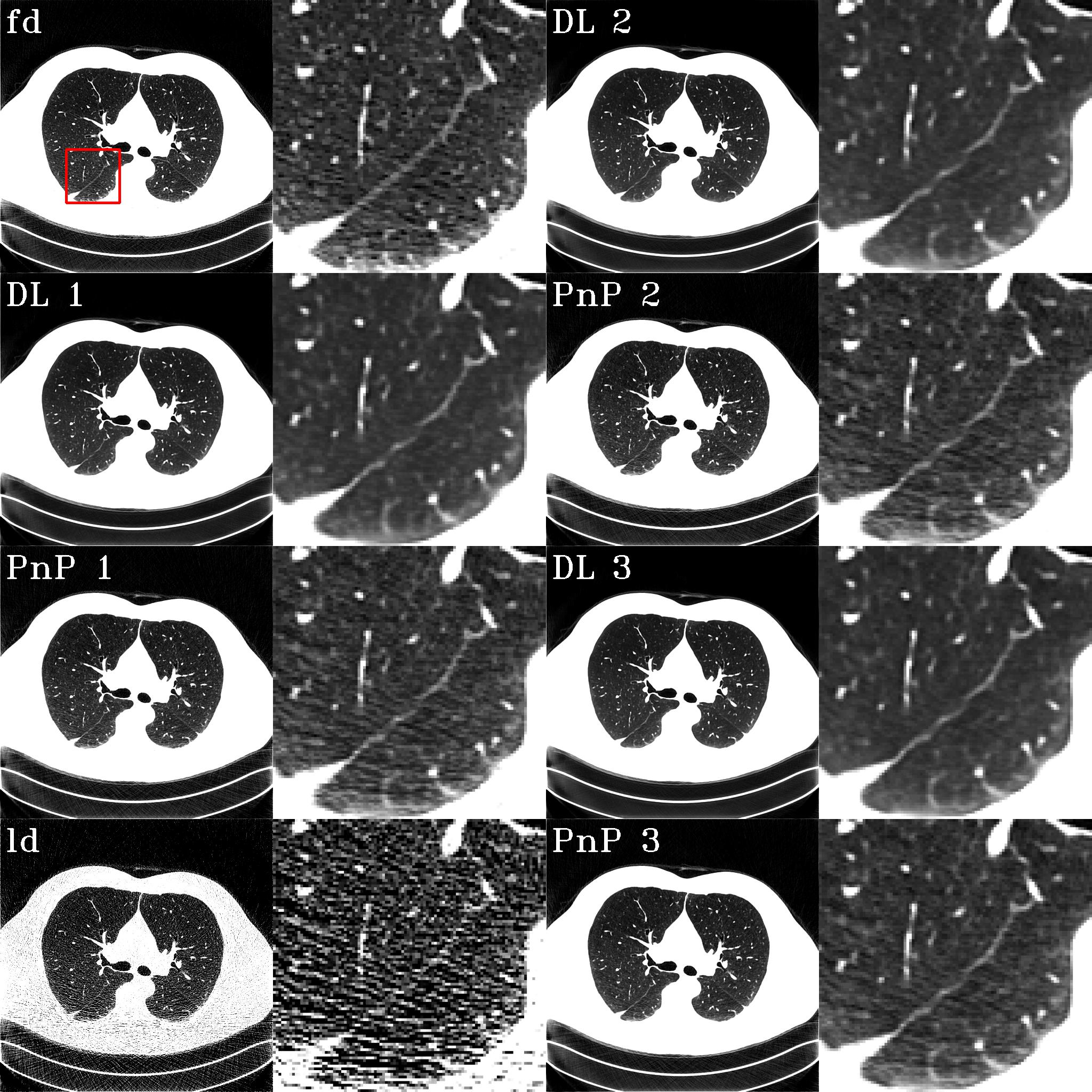}
    \caption{Example images. The red box bounds the enlarged area, which is shown to the right of each image.}
    \label{figure2}
\end{figure}

Fig. \ref{figure2} shows examples of different intermediate results. The conventions in the images are the same as above. The images are shown in a window of [-1000, -267] HU to highlight the lung tissues. One observation is that, as shown in the enlarged area, the DL plugin inevitably smooths the resultant images due to distinct noise patterns of full- and low-dose images. In other words, the noise pattern of full-dose images cannot be predicted, given the noisy low-dose images as input. As a result, to minimize the loss function, the network adopts the strategy to predict a clean image that minimizes the distance expectation over all possible latent full-dose images. While the fine textures are brought back from the sinograms by the optimization process, noise independent of full-dose images is inevitably introduced, resulting in a higher Euclidean distance from the ground-truth full-dose images.

\begin{figure*}
    \centering
    \includegraphics[width=\linewidth]{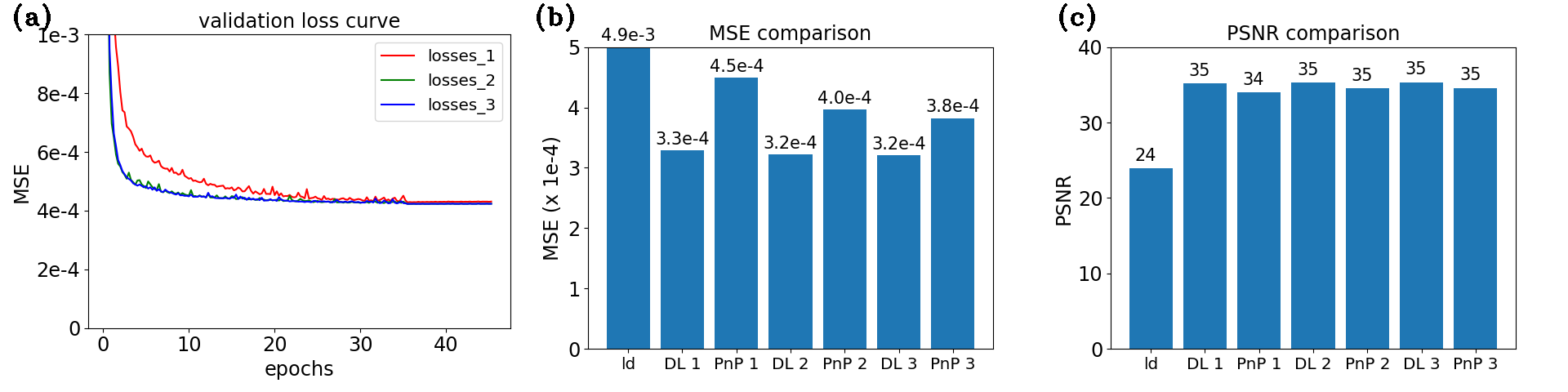}
    \caption{(a) Validation loss curves, (b) MSE, and (c) PSNR on test set.}
    \label{training}
\end{figure*}

\begin{figure}[ht]
    \centering
    \includegraphics[width=\linewidth]{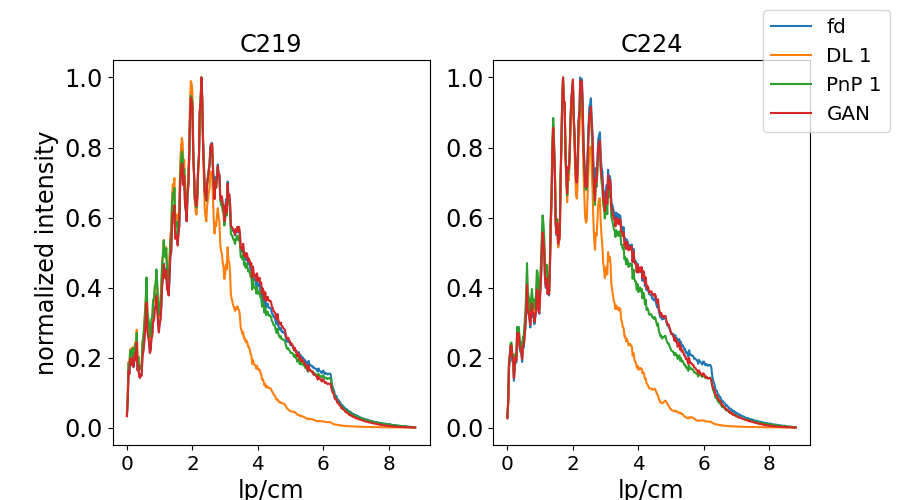}
    \caption{NPS results.}
    \label{NPS}
\end{figure}

\begin{figure}[ht]
    \centering
    \includegraphics[width=\linewidth]{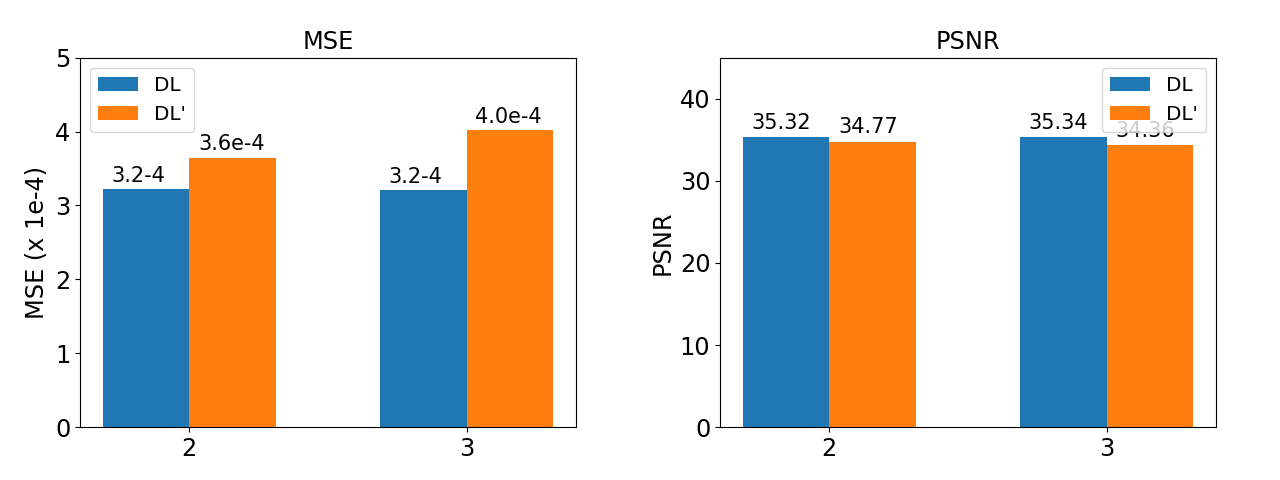}
    \caption{MSE and PSNR comparison between DL images and their DL' counterpart for loops 2 and 3.}
    \label{superior}
\end{figure}

\begin{figure}
    \centering
    \includegraphics[width=\linewidth]{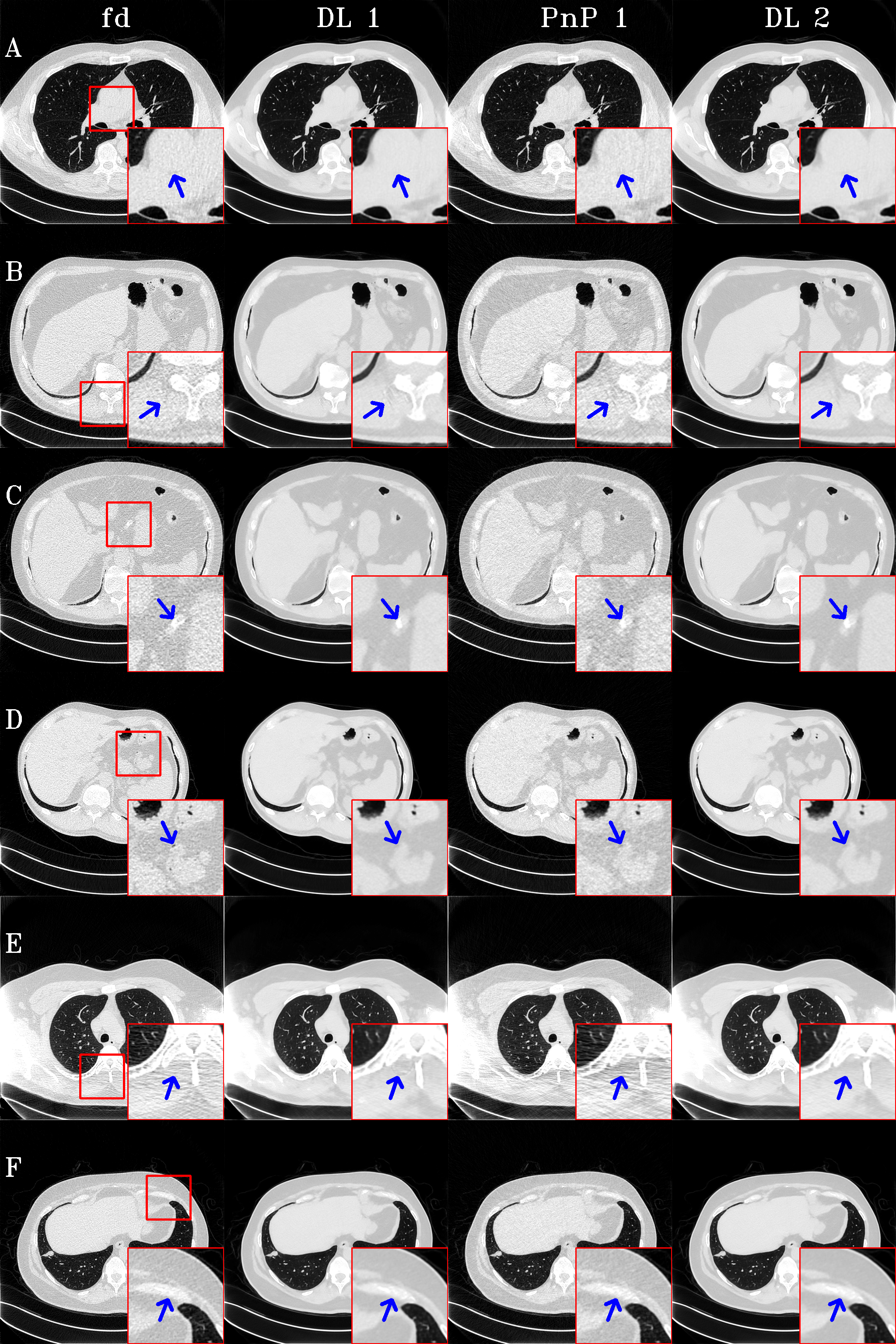}
    \caption{Fine structures are better preserved by PnP, as illustrated by arrows.}
    \label{figure6}
\end{figure}

Fig. \ref{training} (a) shows the validation loss curves of the training of the three loops. The latter two loops converged faster than the first one, which had to work with noisy low-dose images, while the latter two loops worked with relatively clean images. The three curves converged to nearly the same point. The minimum validation MSE values achieved for the three training loops were 4.2918e-4, 4.2324e-4, and 4.2308e-4, respectively. (b) and (c) show the average MSE and PSNR values evaluated on the test set. Values were calculated and averaged over all slices of all patients in the test set.
NPS results for two patients (``C219" and ``C224") are shown in Fig. \ref{NPS}, which manifests that the ``PnP 1" and ``GAN" images have more similar texture properties with full-dose images compared to ``DL 1".
The effectiveness of our loop-specific DL plugin is demonstrated in Fig. \ref{superior}, in which the group using the same DL plugin throughout (DL') shows significantly inferior results than our methods due to DL plugin incompatibility.
Fig. \ref{figure6} provides evidence that the optimization process retrieved additional details from the sinograms, as indicated by the blue arrows. In this figure, the window is set to [-1000, 173] HU to include the muscle and bone areas.

Table \ref{airway} shows airway segmentation results. As expected, the ``PnP 1" images support the most accurate results. On average, segmentation accuracies for ``PnP 1" are 0.6\% and 1.5\% higher than ``DL" and ``GAN" for all airways, and 3.5\% and 4.2 \% for small airways, respectively. For pulmonary lobe segmentation, ``PnP 1" gets the higher scores on 45 out of 50 lung lobes. On average, ``PnP 1" results in 0.5\% and 0.3\% improvements over ``DL 1" and ``GAN", respectively. Wilcoxcon signed-rank test gives a confidence level of over 0.999 that automated segmentation on ``PnP 1" images is more accurate than those of ``DL 1" and ``GAN".

\begin{table}
    \centering
    \begin{tabular}{|c|c|c|c|}
        \hline
        patients & Dice(DL 1) & Dice(GAN) & Dice(PnP 1) \\
        \hline
        C219 & 0.943/0.671 & 0.939/0.674 & \textbf{0.947/0.710}\\ 
        \hline
        C224 & 0.924/0.684 & 0.922/0.706 & \textbf{0.930/0.719}\\ 
        \hline
        C227 & 0.896/0.708 & 0.886/0.704 & \textbf{0.907/0.745}\\ 
        \hline
        C232 & 0.950/0.834 & 0.942/0.809 & \textbf{0.954/0.842}\\ 
        \hline
        C234 & 0.953/0.797 & 0.945/0.775 & \textbf{0.957/0.818}\\ 
        \hline
        C241 & 0.956/0.850 & 0.945/0.826 & \textbf{0.958/0.862}\\ 
        \hline
        C252 & 0.955/0.817 & 0.942/0.791 & \textbf{0.958/0.837}\\ 
        \hline
        C257 & 0.915/0.710 & 0.915/0.715 & \textbf{0.923/0.736}\\ 
        \hline
        C258 & 0.909/0.759 & 0.886/0.742 & \textbf{0.913/0.771}\\ 
        \hline
        C261 & 0.899/0.658 & 0.901/0.694 & \textbf{0.912/0.708}\\ 
        \hline
        Average & 0.930/0.749 & 0.922/0.744 & \textbf{0.936/0.775}\\ 
        \hline
    \end{tabular}
    \caption{airway segmentation results. Two numbers in each box are for the Dice’s coefficients of all airways and small airways with cross section $\leq$ 0.61cm$^2$, respectively. The best performers are shown in bold.}
    \label{airway}
\end{table}

\section{Conclusion}
In this work, a PnP framework with loop-specific plugins was proposed to address domain incompatibility in DL plugins, also eliminating the need of denoising strength tuning. A thorough comparison, including state-of-the-art supervision method, was performed, with perspectives of standard metrics, noise power spectrum, and clinical-relevant segmentation tasks, which demonstrated the advantages of our method in both data fidelity and statistical properties.

\newpage
\section{Acknowledgments}
The study is supported in part by NIH R01CA251874, R01CA255432 and R01CA230278.
\section{Compliance with Ethical Standards}
This is a computational study involving only publicly available datasets. No ethical approval was required.
\bibliographystyle{IEEEbib}
\bibliography{Template_ISBI_latex}

\end{document}